\documentclass[aps,prl,reprint,twocolumn,showpacs,floatfix,superscriptaddress]{revtex4-2}

\usepackage{amssymb,amsmath,amstext}                
\usepackage{graphicx}                                               

\usepackage{fp}
\newcount\boxheight
\newcount\boxwidth
\newcommand\testaspect[1]{%
  \setbox0=\hbox{#1}%
  \boxheight=\ht0\relax%
  \boxwidth=\wd0\relax%
  \FPdiv\theaspect{\the\boxheight}{\the\boxwidth}%
  \copy0%
}

\usepackage{epstopdf}                                               
\usepackage{color}                                                     
\usepackage{bm}                                                        
\usepackage{comment}     
\usepackage{appendix}                                              
\usepackage[utf8]{inputenc}
\usepackage{bbold}
\usepackage{bbm}
\usepackage{latexsym}
\usepackage[T1]{fontenc}
\usepackage{xcolor}
\usepackage{braket}
\definecolor{lblue} {RGB}{51,71,158}
\usepackage[colorlinks=true,citecolor=blue,linkcolor=blue,urlcolor=lblue]{hyperref}

\usepackage[normalem]{ulem}

\definecolor{darkgreen}{rgb}{0.13, 0.55, 0.13}

\def\beq{\begin{equation}}
\def\eeq{\end{equation}}

\usepackage{dcolumn}
\usepackage{sidecap}
\usepackage[caption=false]{subfig}
\usepackage{bm}

\begin{document}

\title{Scar States in Deconfined $\mathbb{Z}_2$ Lattice Gauge Theories}

\author{Adith Sai Aramthottil} 
\affiliation{Instytut Fizyki Teoretycznej, 
Uniwersytet Jagiello\'nski,  \L{}ojasiewicza 11, PL-30-348 Krak\'ow, Poland}
\author{Utso Bhattacharya}
\affiliation{ICFO-Institut de Ci\`encies Fot\`oniques, The Barcelona Institute of Science and Technology, Av. Carl Friedrich
Gauss 3, 08860 Castelldefels (Barcelona), Spain}
\author{Daniel Gonz\'alez-Cuadra}
\affiliation{ICFO-Institut de Ci\`encies Fot\`oniques, The Barcelona Institute of Science and Technology, Av. Carl Friedrich
Gauss 3, 08860 Castelldefels (Barcelona), Spain}
\affiliation{Center for Quantum Physics, University of Innsbruck, 6020 Innsbruck, Austria}
\affiliation{Institute for Quantum Optics and Quantum Information of the Austrian Academy of Sciences, 6020 Innsbruck, Austria}
\author{Maciej Lewenstein} 
\affiliation{ICFO-Institut de Ci\`encies Fot\`oniques, The Barcelona Institute of Science and Technology, Av. Carl Friedrich
Gauss 3, 08860 Castelldefels (Barcelona), Spain}
\affiliation{ICREA, Passeig Lluis Companys 23, 08010 Barcelona, Spain}
\author{Luca Barbiero}
\affiliation{Institute for Condensed Matter Physics and Complex Systems,
DISAT, Politecnico di Torino, I-10129 Torino, Italy}
\affiliation{ICFO-Institut de Ci\`encies Fot\`oniques, The Barcelona Institute of Science and Technology, Av. Carl Friedrich
Gauss 3, 08860 Castelldefels (Barcelona), Spain}
\author{Jakub Zakrzewski} 
\affiliation{Instytut Fizyki Teoretycznej, 
Uniwersytet Jagiello\'nski,  \L{}ojasiewicza 11, PL-30-348 Krak\'ow, Poland}
\affiliation{Mark Kac Complex Systems Research Center, Uniwersytet Jagiello{\'n}ski, Krak{\'o}w, Poland}

\date{\today}

\begin{abstract}
The weak ergodicity breaking induced by quantum many-body scars (QMBS)
represents an intriguing concept that has received great attention in recent years due to its relation to unusual non-equilibrium behaviour. Here we reveal that this phenomenon can occur in a previously unexplored regime of a lattice gauge theory, where QMBS emerge due to the presence of an extensive number of local constraints. In particular, by analyzing the gauged Kitaev model, we provide an example where QMBS appear in a { regime where charges are} deconfined.
By means of both numerical and analytical approaches, we find a variety of scarred states 
far away from the { regime where the model is} integrable. The presence of these states is revealed both by tracing them directly from the analytically reachable limit, as well as by
quantum quenches showing persistent oscillations for specific initial states.
\end{abstract}

\maketitle

\paragraph*{Introduction.--}
The thermalization properties of isolated quantum systems are currently under intensive investigation in different areas of modern quantum physics \cite{Rigol08,Polkovnikov11,Gogolin2016,Vidmar16}. In this context, a huge attention has been recently devoted towards the study  
of a large variety of Hamiltonians where the ergodicity is weakly broken \cite{Bernien17,Shiraishi17,Turner18,Turner18q,Moudgalya18,Moudgalya18en,Iadecola19, Iadecola19a,Iadecola20,Serbyn20,Lee20,Surace20,Magnifico20,Chanda20,Zhao20, Papic21,Moudgalya21,Surace21q,Zhao21,Szoldra22}.  In particular, in such models
quenches from carefully designed initial states reveal 
persistent many-body revivals that apparently contradict ergodicity~\cite{Bernien17,Bluvstein21}.
The reason for such an absence of thermalization turns out to be the presence 
of specific eigenstates called quantum many-body scars (QMBS) 
characterized, in particular, by 
a subvolume entanglement law \cite{Shiraishi17,Turner18, Turner18q,Moudgalya18,Moudgalya18en,Iadecola19,Iadecola19a,Iadecola20,Serbyn20,Moudgalya21,Turner21,Szoldra22}.
 It is worth noting that similar regular states have been previously identified \cite{Heller84} for  chaotic  quantum billiards in relation to semiclassical periodic orbit quantization \cite{Gutzwiller71,Bogomolny88}. Moreover, the concept of scarred symmetry, closely related to QMBS, has been introduced in studies of hydrogen atom in a strong magnetic field \cite{Delande87}.\\
 \begin{figure}
\centering
	\includegraphics[width=0.9\linewidth]{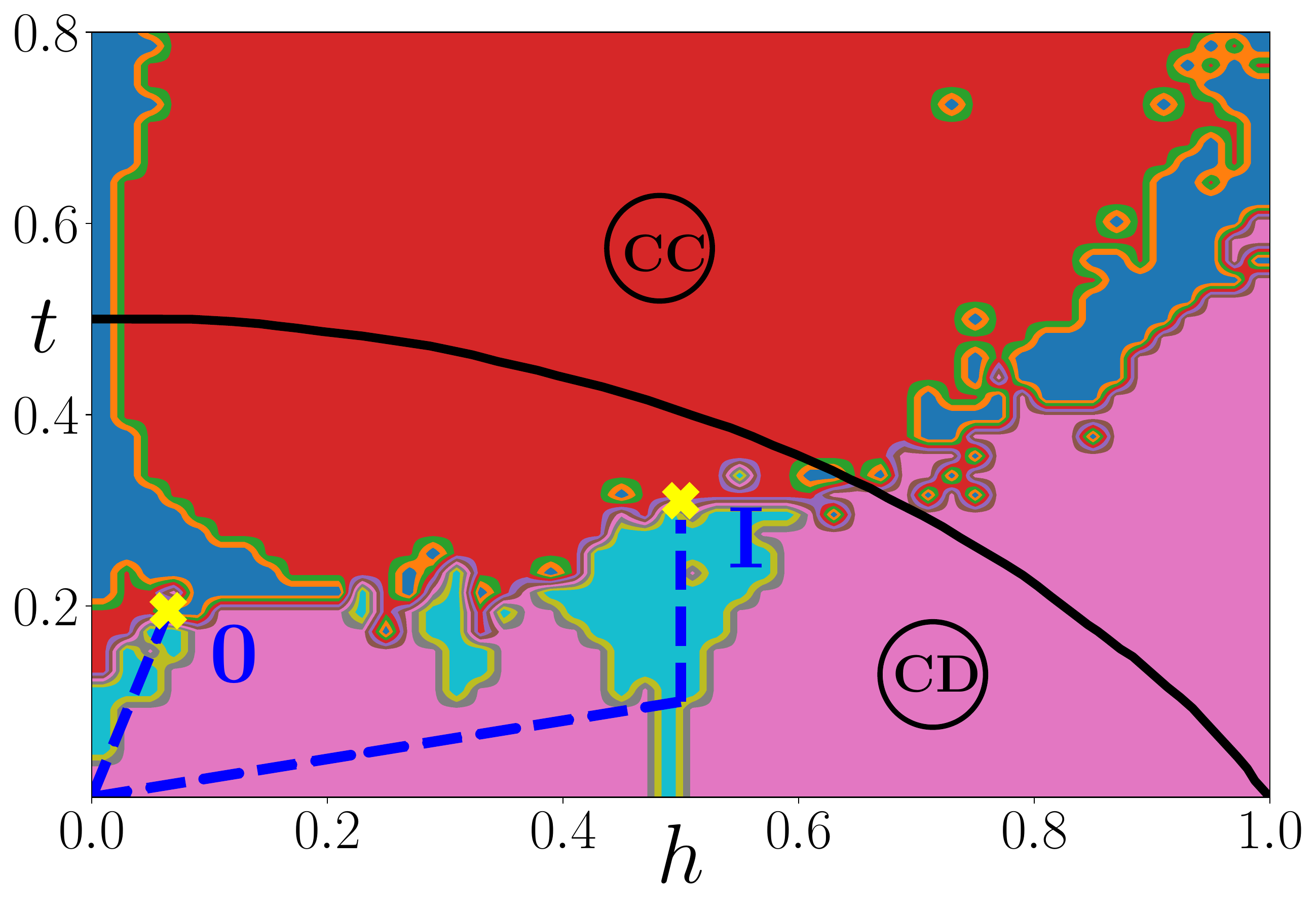}
	\caption{Properties of the  Ising model \eqref{eq:Isi} in $t-h$ parameter space for $L=16$ and $\mu=1$. QMBS may exist in chaotic regions with mean gap ratio $r>0.5$ and low entanglement entropy $S<0.5$ with respect to the GOE value. The corresponding disconnected blue region lies entirely in charge deconfined (CD) regime. Charge confinement (CC) region lies mainly within the regime of chaotic, $r>0.5$, but with lowest entropy states, $S>0.5$, thus with no chance for QMBS. Pink color marks the mixed dynamics regime, $r<0.5$, $S<0.5$; in dark blue region while $r<0.5$, $S>0.5$. `0' and `I' denote paths along which QMBS states are followed using level dynamics.}
	\label{fig:contour}
\end{figure}
With the advent of a new generation of cold-atom quantum simulators~\cite{Jaksch05,Lewenstein07,Gross17,Schafer20,Browaeys20}, the weak ergodicity breaking manifested by QMBS has been experimentally detected in constrained spin ~\cite{Bernien17} and bosonic ~\cite{Su2022} models. Crucially, both realizations can be viewed as lattice gauge theories (LGTs) where the energy constraints are induced by the Gauss law fixing the relation between gauge and charge variables. Motivated by the recent progress achieved in the last years in implementing LGTs in quantum simulators \cite{Martinez2016,Schweizer2019,Gorg2019,Mil2020,Yang20,Yang20N},  an impressive theoretical effort has been devoted towards a better understanding of simple gauge-invariant theories~\cite{Zohar11,Dalmonte2016,Assaad16,Gonzalez17,Gazit2017,Gazit18, Barbiero2019,Borla2020,Magnifico20,Chanda20,Gonzalez20,Banuls20,Halimeh20,Halimeh21,VanDamme21,Kebric2021,Aidelsburger22}. In this direction, the connection between Gauss law and QMBS has recently gained attention both in $U(1)$ \cite{Surace20,Banerjee21} and $\mathbb{Z}_2$ \cite{Iadecola20} LGTs. Here indeed, the weak ergodicity breaking associated with the slow oscillatory dynamics can be interpreted as a \textit{string inversion} phenomenon. Crucially, it has to be underlined that all LGTs, where QMBS have been identified, are characterized by charge confinement. This regime implies that only particle-antiparticle bound states exist and therefore charges can  be observed in composite
structures only. These effective pairs,  together with an emergent new symmetry, generate the slowdown of quantum dynamics and have been shown to be intricately connected to the presence of QMBS \cite{Chen21}.\\
Thus it seems natural to wonder whether confinement is a prerequisite to observe QMBS in LGTs. In this Letter, we tackle this question by investigating the recently introduced gauged 1D Kitaev model~\cite{Lerose2020,Borla21,Surace20}, whose ground state displays a confined phase as well as a regime where charges are not bounded in pairs, thus describing a deconfined phase. 
We study its low-entanglement states and show that QMBS are present in the ergodic deconfined phase and are absent in the ergodic confined regime of the model. 
 Importantly, we are able to continuously track  QMBS down from the analytic prediction valid in the quasi-integrable regime  and therefore
provide their partial classification.

\paragraph*{Model and observables.--}
The Hamiltonian of the $p$-wave superconducting Kitaev chain minimally coupled to a $\mathbb{Z}_2$ gauge field 
introduced in \cite{Lerose2020,Borla21,Surace20} reads
\begin{eqnarray}
H &=& -t\sum_j\left(c^\dagger_j-c_j\right)\sigma^z_{j+1/2}\left(c^\dagger_{j+1}+c_{j+1}\right)\nonumber\\
&~&-\mu\sum_j\left(c^\dagger_jc_j-\frac{1}{2}\right)-h\sum_j\sigma^x_{j+1/2}.
\label{eq:kit}
\end{eqnarray} 
Here, $c^\dagger_j$ ($c_j$) denotes the  fermionic creation (annihilation) operator and $t$ describes the tunneling and pair production/annihilation processes mediated by the $\mathbb{Z}_2$ gauge field $\sigma^z_{j+1/2}$ defined on the links between the nearest neighbor sites. Fluctuations in the gauge field are induced by the electric field  $\sigma^x_{j+1/2}$ of strength $h$ and the number of fermions is fixed by the chemical potential $\mu$. {Here $\sigma^i_{j+1/2}$ stand for standard Pauli matrices.}

As required in LGTs, the model (\ref{eq:kit}) is invariant under the local gauge transformation generated by the Gauss operator $G_j = \sigma^x_{j-1/2}(-1)^{n_j}\sigma^x_{j+1/2}$ where $[H,G_j]=0$ and $[G_i,G_j]=0$. Therefore, the physical states are those that satisfy the Gauss law, $G_j\ket{\psi} =\pm \ket{\psi}$, for all $j$ sites~\cite{Kogut1979}.

The Hamiltonian (\ref{eq:kit}) can be written in terms of gauge-invariant {transformed Pauli operators}:
\begin{eqnarray}
X_{i+\frac{1}{2}} &=& \sigma^x_{i+\frac{1}{2}}\\
Y_{i+\frac{1}{2}} &=& \left(c^\dagger_i-c_i\right)\sigma^y_{i+\frac{1}{2}}\left(c^\dagger_{i+1}+c_{i+1}\right)\\
Z_{i+\frac{1}{2}} &=& \left(c^\dagger_i-c_i\right)\sigma^z_{i+\frac{1}{2}}\left(c^\dagger_{i+1}+c_{i+1}\right).
\end{eqnarray}
Upon this transformation and with periodic boundary conditions, the gauged Kitaev model corresponds to the quantum Ising model with 
both transverse and longitudinal fields
\begin{equation}
	H = \sum_{i=1}^{L} \frac{\mu}{2} Z_i Z_{i+1} - t X_i   - h Z_i.
	\label{eq:Isi}
\end{equation}
Notice that the model is now defined on a dual lattice, where the index $i$ corresponds to the links of the original model~\eqref{eq:kit}.

For $\mu>0$, the phase diagram of the model (\ref{eq:Isi}), as shown in Fig.~\ref{fig:contour},  is characterized by the presence of both an antiferromagnetic (AFM) order  and a paramagnetic (PM) phase depending on parameters $t$ and $h$ ~\cite{Ovchinnikov03}. The AFM phase turns out to be of great interest, since it supports domain-wall excitations associated with the antiferromagnetic order caused by the spontaneous breaking of the $\mathbb{Z}_2$ Ising and translational symmetries. In 1D these last two features imply  
domain wall  deconfinement. 
Therefore, in the gauged Kitaev Hamiltonian, (\ref{eq:kit}), such an AFM order corresponds to charge deconfinement (CD), where fermions are free to expand without any string tension.  On the other hand, the PM regime corresponds to a phase characterized by charge confinement (CC), where fermions appear only as bound pairs.  

In the limit $t\ll \mu$ it is possible to
perform  Schrieffer-Wolff transformation (in agreement with \cite{Lerose2020}, for a higher order expansion see \cite{Moroz21}), with
\begin{eqnarray}
S &=& \frac{-it}{2\mu}\sum_j\Bigg\{\left(\frac{1+Z_{j-1}}{2}\right)Y_j\left(\frac{1+Z_{j+1}}{2}\right)\nonumber\\
&~~~~~~~&-\left(\frac{1-Z_{j-1}}{2}\right)Y_j\left(\frac{1-Z_{j+1}}{2}\right)\Bigg\},
\end{eqnarray}
 to obtain an effective Hamiltonian $H_{\rm eff} = e^S H e^{-S} = H + [S,H] + \mathcal{O}(t^2)$. Notice that $S$ is chosen in such a way that the new terms 
 commute with the unperturbed part of the Hamiltonian $\sum_{i=1}^{L} \frac{\mu}{2} Z_i Z_{i+1}$ to the leading order in the expansion, thus preserving its block-diagonal structure while also including all virtual processes within each sector. 
The derived effective Hamiltonian,
\begin{equation}
    H_{eff} = \sum_{i=1}^{L} \frac{\mu}{2} Z_i Z_{i+1} - h Z_i - \frac{t}{2}\left(X_i - Z_{i-1}X_iZ_{i+1}\right)
    \label{eq:iadec}
\end{equation}
turns out to be the well-known model discussed in detail in \cite{Iadecola20}, where two towers of QMBS have been identified. The latter are of the form
\begin{equation}
	\vert S_n^{k} \rangle = \frac{1}{n!\sqrt{\mathcal{N}(L,n)}}((\mathcal{Q}^{k})^\dagger)^n\vert \Omega^{k} \rangle,
	\label{scar}
\end{equation}  
with $k=1,2$, $\vert \Omega^1 \rangle=\vert 0\cdots 0 \rangle$, $\vert \Omega^2 \rangle=\vert 1\cdots 1 \rangle$ and {$(\mathcal{Q}^{k})^\dagger=\sum_{i=i_1}^{i_L}(-1)^iP_{i-1}^{k}(X_i+(-1)^kY_i) P_{i+1}^{k}$, with  projection operators $P_i^{k}=(1+(-1)^kZ_i)/2$.}
Such QBMS describe $n$-magnon and $n$-antimagnons excitations for $k=1$ and $k=2$, respectively \cite{Iadecola20}.

\begin{figure}%
	\includegraphics[width=\linewidth]{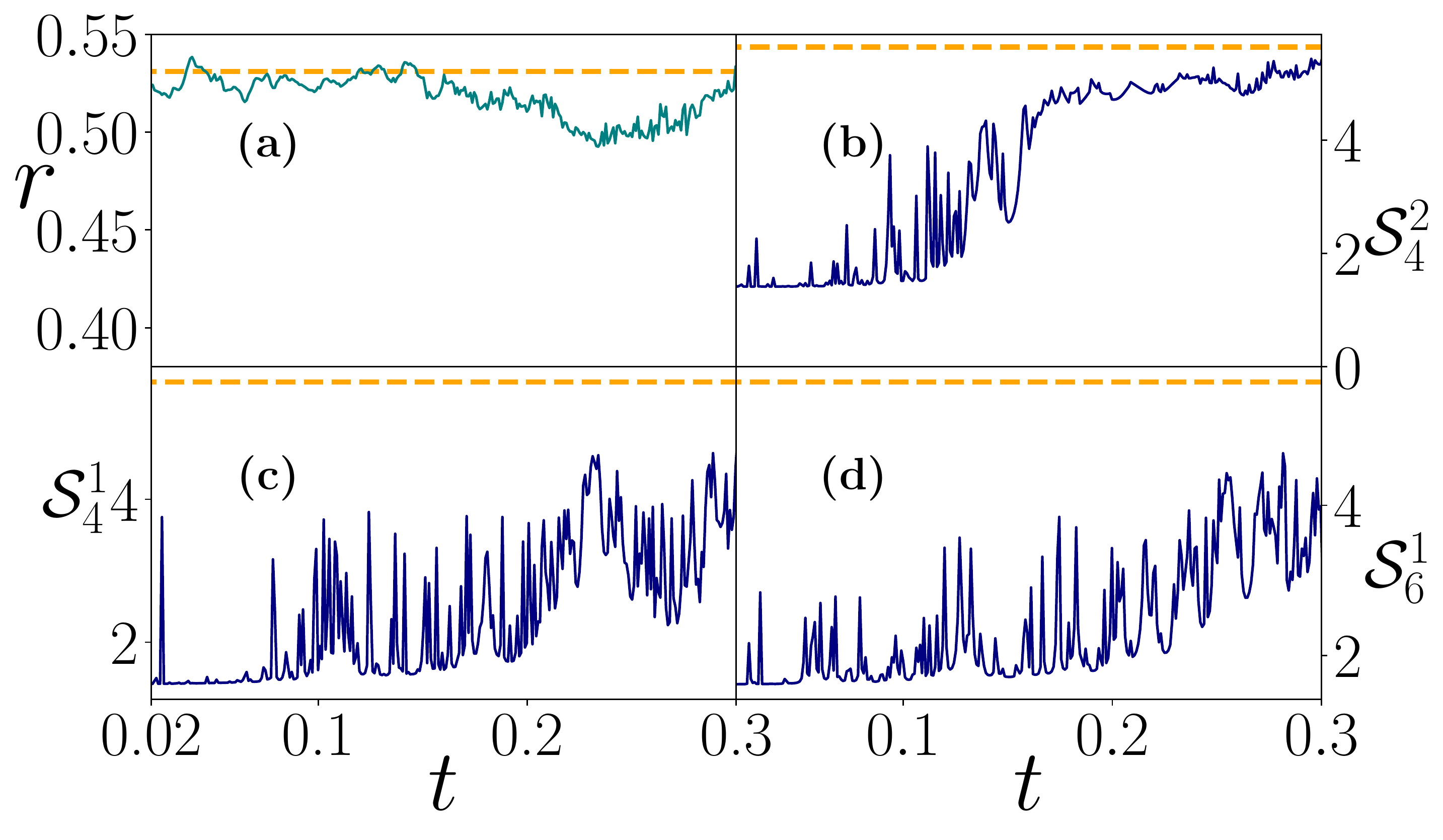}
	\caption{The gap ratio $r$, (a) and half-chain entanglement entropy($\mathcal{S}$) along path 0 for  (b)-- $S^2_4$ state form antimagnon family; (c) -- $S^1_4$ and (d) -- $S^1_6$ from the magnon family, all for $L=18$. States may be followed despite narrow avoided crossings indicated by spikes of the entanglement entropy.  The antimagnon  state looses
	its  low EE feature around $t=0.14$ while magnon-like QMBS may be followed up to $t\approx0.2$.}
	\label{fig:new2}
\end{figure}

As already pointed out, the Schrieffer-Wolff transformation described above links the models~\eqref{eq:Isi} and ~\eqref{eq:kit} with the effective spin model~\eqref{eq:iadec} only for $t\ll \mu $. 
In this limit, the states given by Eq.~\eqref{scar} 
become true QMBS of Hamiltonian \eqref{eq:iadec}, as it was studied in Ref.~\cite{Iadecola20}. It appears natural to wonder whether,   
following the states \eqref{scar} in the parameter space, it is possible to also find scarred states in the gauged Kitaev chain beyond the $t\ll \mu $ limit.

\paragraph*{Level dynamics.--}
In order to investigate this point, we first determine in which 
regime of parameters the model \eqref{eq:Isi} can be considered as ergodic, and thus where the regular states may be called QMBS.  Ergodicity may be revealed by the adjacent mean gap ratio $r$ \cite{Oganesyan07} between subsequent level spacings $\Delta_i$, 
\begin{equation}
	r_i=\frac{min\lbrace\Delta_i,\Delta_{i+1}\rbrace}{max\lbrace\Delta_i,\Delta_{i+1}\rbrace}
\end{equation}
where $r \simeq 0.531$ corresponds to the fully
ergodic regime (as described by the Gaussian Orthogonal Ensemble (GOE) - \cite{Haakebook}) and $r \simeq 0.386$ indicates the quasi-integrable regime \cite{Oganesyan07,Atas13}. As shown in Fig.~\ref{fig:contour}, our model shows strong indications of near-integrable behavior for $h \ll \mu$ or $t \ll \mu$. Outside this region, the model is expected to be non-integrable.  The next step is then to identify  the region where the entanglement entropy (EE) of some eigenstates is well below the GOE estimate,
as QMBS should have low, subvolume entropy. The half-chain EE is defined in a standard way as  $\mathcal{S}=-Tr[\rho_{L/2}\ln{\rho_{L/2}}]$ \cite{suppl}. We compare it to the typical GOE value for a given system size. Once the two lowest EE states have a relative EE sufficiently low (say half of the GOE value), there is the chance that those states are indeed scarred. Fig.~\ref{fig:contour} shows by a light blue color the domains where both $r>0.5$ and states with a sufficiently low EE exist. 

With the suspected regions identified,  we start the analysis with $\vert S_n^{k}\rangle$ states  for the Hamiltonian \eqref{eq:Isi} with parameters $t, h \ll \mu$, e.g. 
for  $t_0=h_0=0.001$ (we set $\mu=1$ in the following), with the aim of tracking such initial states $\vert S^k_n(t_0,h_0)\rangle$ following their possible deformations induced by making small changes in the parameters,  $t =  t_0+\delta t$ and $h = h_0+\delta h$. As examples we consider paths indicated by `0' and `I' in Fig.~\ref{fig:contour}.

Every time we update the parameters, we diagonalize the Hamiltonian in the symmetry sector with momentum $p=0$ and parity $+1$ (we consider $n$ even only) and find the new candidate $\vert E_n(t,h)\rangle$
by maximizing the overlap $O=|\langle E_n(t,h)\vert S^k_n)\rangle|$. For an isolated level, the new state is accepted, $\vert S^k_n\rangle$ is updated, and we repeat the procedure. Special care is taken to diabatically  cross narrow avoided crossings \cite{suppl}.

\begin{figure}%
	\includegraphics[width=\linewidth]{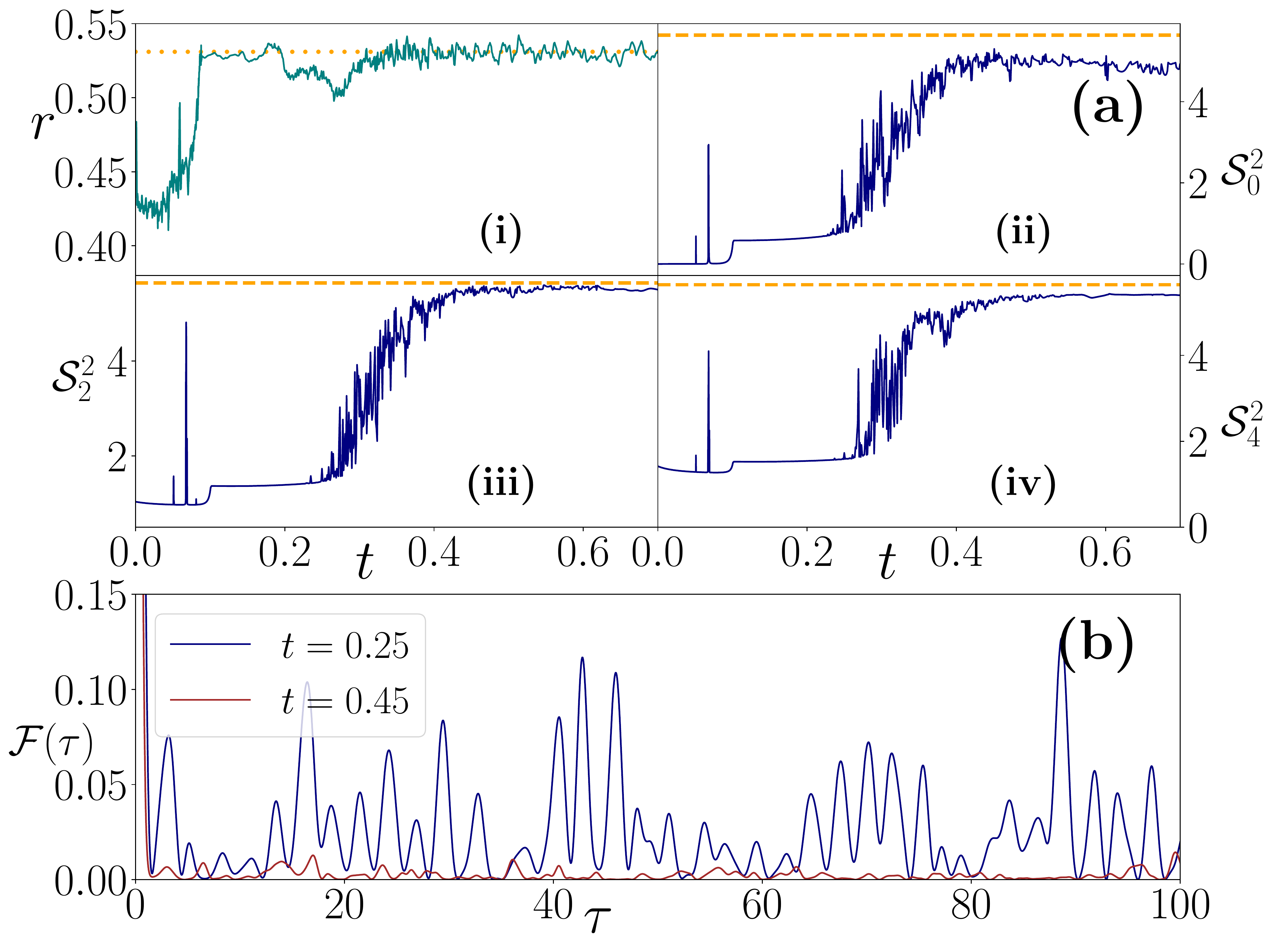}
	\caption{Panel (a) is like  Fig.~\ref{fig:new2} but for path I and $L=18$. Along this path only antimagnon excitations may be followed while magnons appear at the edges of the energy spectrum only. (b) Time dependence of the fidelity relative to an initial state $\vert \psi(0) \rangle$ and evolved with (\ref{eq:Isi}) at $h=0.5$ and different values of $t$ for $L=16$ ($\mu=1$).}
	\label{fig:new3}
\end{figure}

Figure~\ref{fig:new2} shows tracking the QMBS states identified as $\vert S^k_n(t,h)\rangle$ along path 0. Both magnon and antimagnon excitations, as given by \eqref{scar}, may be followed. This path lies almost fully in the chaotic regime as revealed by $r$ statistics, so observed low EE states may be truly considered as QMBS,
observed, let us stress, deep in the deconfined region.

The path indicated as I also remains in the deconfined phase, extending to large values of $h$ - c.f. Fig.~\ref{fig:contour}.   Levels are followed through regular, weakly perturbed region, along the straight line $t=0.2h$ up to $h=0.5$. Then path I turns upwards, staying in the chaotic region up to the point $(t, h) = (0.3, 0.5)$.  Along this vertical part,  the tracked low EE states, shown in Fig.~\ref{fig:new3}(a), are truly the QMBS embedded in the chaotic spectrum. As seen from the EE values, the scarred character of the followed states is slowly lost due to, as verified, \cite{suppl} various avoided crossings. 

As discussed in Introduction, the presence of QMBS can be further revealed by the persistent time oscillations of an out-of-equilibrium configuration. To reveal this aspect, we prepare an initial state given by equal superposition of two states of the form \eqref{scar} for small $t,h$ values, $\vert \psi(0) \rangle =\frac{1}{\sqrt{2}}( \vert S_0^{2}\rangle + \vert S_2^{2}\rangle)$. After this initialization, we let this state evolve with Hamiltonian \eqref{eq:Isi} and we calculate the fidelity $\mathcal{F}(\tau)=\vert \langle \psi(0)\vert \psi(\tau)\rangle \vert^2 $, where $\tau$ is the time (in units of inverse $\mu$). At $t=0.25$, we observfor other choices of finitee persistent oscillations of $\mathcal{F}(\tau)$ as expected for QMBS states, and this behavior persists until the end of path I - compare Fig.~\ref{fig:new3}(b). Beyond this limit ($t > 0.3$), low entanglement states disappear and $\mathcal{F}(\tau)$ shows irregular oscillations around the mean value of about $1/L^2$, as expected for thermal states.

\begin{figure}%
	\includegraphics[width=\linewidth]{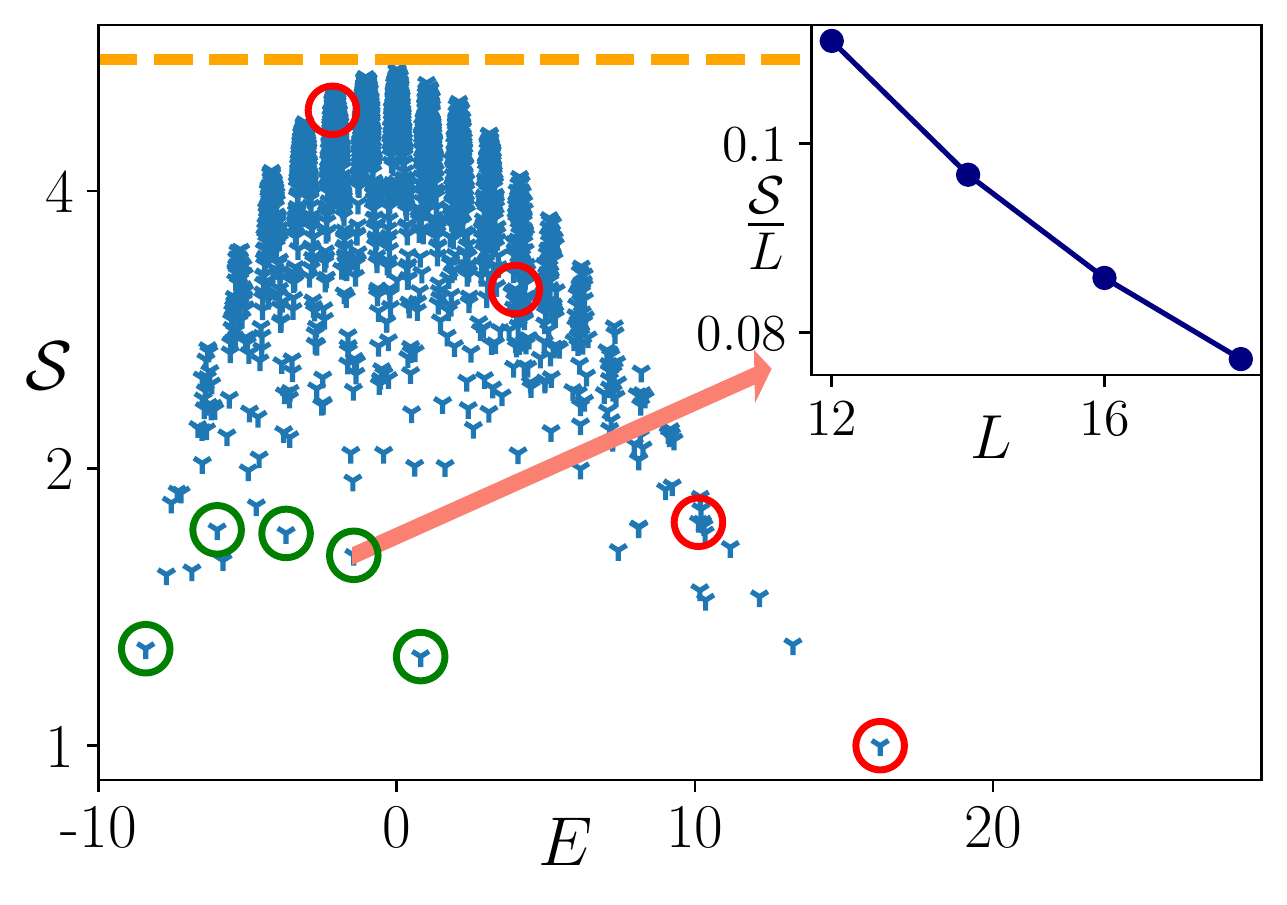}
	\caption{The half-chain entanglement entropy($\mathcal{S}$)  of all the eigenstates at  $t=0.2$, $h=0.5$ for $L=16$. The orange dashed line gives the $\mathcal{S}_{RMT}$ value. Circles denote different QMBS obtained via our tracking procedure. Green circles denote antimagnon-like family $S^2_n$ for $n=0,2,4,6,8$  while red circles  magnon-like states, $S^1_n$ with $n=0,..,6$ counting from the right hand side. \textit{Inset:} The half-chain Entanglement Entropy divided by system size ($\frac{\mathcal{S}}{L}$) 
	for $S^2_2$ state showing its sub-volume property as expected for QMBS.}
	\label{fig:pathIb_new}
\end{figure}

Figure~\ref{fig:pathIb_new} visualizes the presence of QMBS by showing the value of $\mathcal{S}$ for all eigenstates at $t = 0.2$ and $h = 0.5$, where the system is in the weak ergodic regime.
The entropy still reveals a finger-like structure that indicates the existence of hidden, unidentified symmetries. 
The states enclosed in circles are those tracked up from the near-integrable limit. The members of the antimagnon-like family $S^2_k$, denoted by green circles, have very low entanglement entropies as compared to other states of similar energy. They are thus truly QMBS. {The inset reveals a sub-volume scaling of the entanglement entropy of the $S^2_2$ state.} The magnon excitations, on the other hand, dissolve among other states (in the high density of states region).

In summary, motivated by recent predictions of finding QMBS states in a confined regime \cite{James19,Chen21} of LGTs, 
we investigated the thermalization properties of  the gauged Kitaev chain.
We observe that deep in the confined phase even states with the smallest entanglement entropy have relatively large, volume-law values indicating a lack of QMBS. On the other hand, for relatively large values of the electric field, some states reveal low entanglement entropy. We can identify these states by following them from the very small $t,h$ values for which analytic predictions are available \cite{Iadecola20}. 
Such states result to be true QMBS and are found  
uniquely in the deconfined phase, which turns out to be a weakly ergodic regime characterized by the GOE-like mean gap ratio. The presence of QMBS has been further verified by studying their time dynamics.  By building  an  initial state given by the superposition of two QMBS, the fidelity of the state shows pronounced oscillations with no sign of thermalization.
The adiabatic following of the states breaks down at the end points of our chosen paths due to energy-level mixing, where interestingly QMBS also disappear. In conclusion, our results unambiguously reveal that QMBS occur even in deconfined regimes of LGTs, thus paving the way toward a deeper understanding of the connection between lack of thermalization and local symmetries.

\begin{acknowledgments}
We thank  A. Bohrdt, U. Borla, F. Grusdt, J.C. Halimeh, P. Hauke. M. Kebric, S. Moroz and P. Sierant for discussions on this manuscript and related projects.
The numerical computations have been possible thanks to   PL-Grid Infrastructure.
The works of A.S.A. and J.Z have been realized within the Opus grant
 2019/35/B/ST2/00034, financed by National Science Centre (Poland).
 M.L. acknowledges support from ERC AdG NOQIA, State Research Agency AEI (``Severo Ochoa'' Center of Excellence CEX2019-000910-S, Plan National FIDEUA PID2019-106901GB-I00/10.13039 / 501100011033, FPI, QUANTERA MAQS PCI2019-111828-2 / 10.13039/501100011033), Fundacio Privada Cellex, Fundacio Mir-Puig, Generalitat de Catalunya (AGAUR Grant No. 2017 SGR 1341, CERCA program, QuantumCAT U16-011424, co-funded by ERDF Operational Program of Catalonia 2014-2020), EU Horizon 
 2020 FET-OPEN OPTOLogic (Grant No. 899794), and the National Science Centre, Poland (Symfonia Grant No. 2016/20/W/ST4/00314), Marie Sk{\l}odowska-Curie grant STRETCH No. 101029393, La Caixa Junior Leaders fellowships (ID100010434),  and EU Horizon 2020 under Marie Sk{\l}odowska-Curie grant agreement No 847648 (LCF/BQ/PI19/11690013, LCF/BQ/PI20/11760031,  LCF/BQ/PR20/11770012). D.G.-C. is supported by the Simons Collaboration on Ultra-Quantum Matter, which is a grant from the Simons Foundation (651440, P.Z.).
\end{acknowledgments}

%


\newcommand{\snum}{S}

\renewcommand{\theequation}{\snum.\arabic{equation}}
\renewcommand{\thefigure}{S\arabic{figure}}
\renewcommand{\bibnumfmt}[1]{[S#1]}
\setcounter{equation}{0}
\setcounter{figure}{0}

 \pagebreak

 \section{Supplementary material for Scar States in Deconfined  Lattice Gauge Theories }
 \label{appendix1}

\subsection{Entanglement Entropy}

We use the half-chain entanglement entropy, $\mathcal{S}$, as a useful measure of eigenstates properties. 
The half-chain entanglement entropy is defined as the von Neumann entropy of the reduced density matrix $\rho_{L/2}$ 
\begin{equation}
	\mathcal{S}=-Tr[\rho_{L/2}\ln{\rho_{L/2}}].
\end{equation}
where $\rho_{L/2} = Tr_{1,\cdots,L/2} \vert \psi \rangle \langle \psi \vert $ is obtained by tracing
out half of the system for eigenstate $ \vert \psi \rangle $. The typical state in the ergodic regime obeys RMT prediction given by $\mathcal{S}_{RMT}=(L/2)\ln(2)+(1/2+\ln(1/2))/2-1/2$ \cite{Vidmar17,Huang19b,Huang21}. A substantial difference from $\mathcal{S}_{RMT}$ for states in the middle of the spectrum can be taken as a signature of nonergodic states, although to really observe a sub-volume property of the entanglement
entropy for a given state one has to study the entropy dependence on the system size (see the inset in  Fig.~\ref{fig:pathIb_new} in the main text.
\begin{figure}%
	\includegraphics[width=0.9\linewidth]{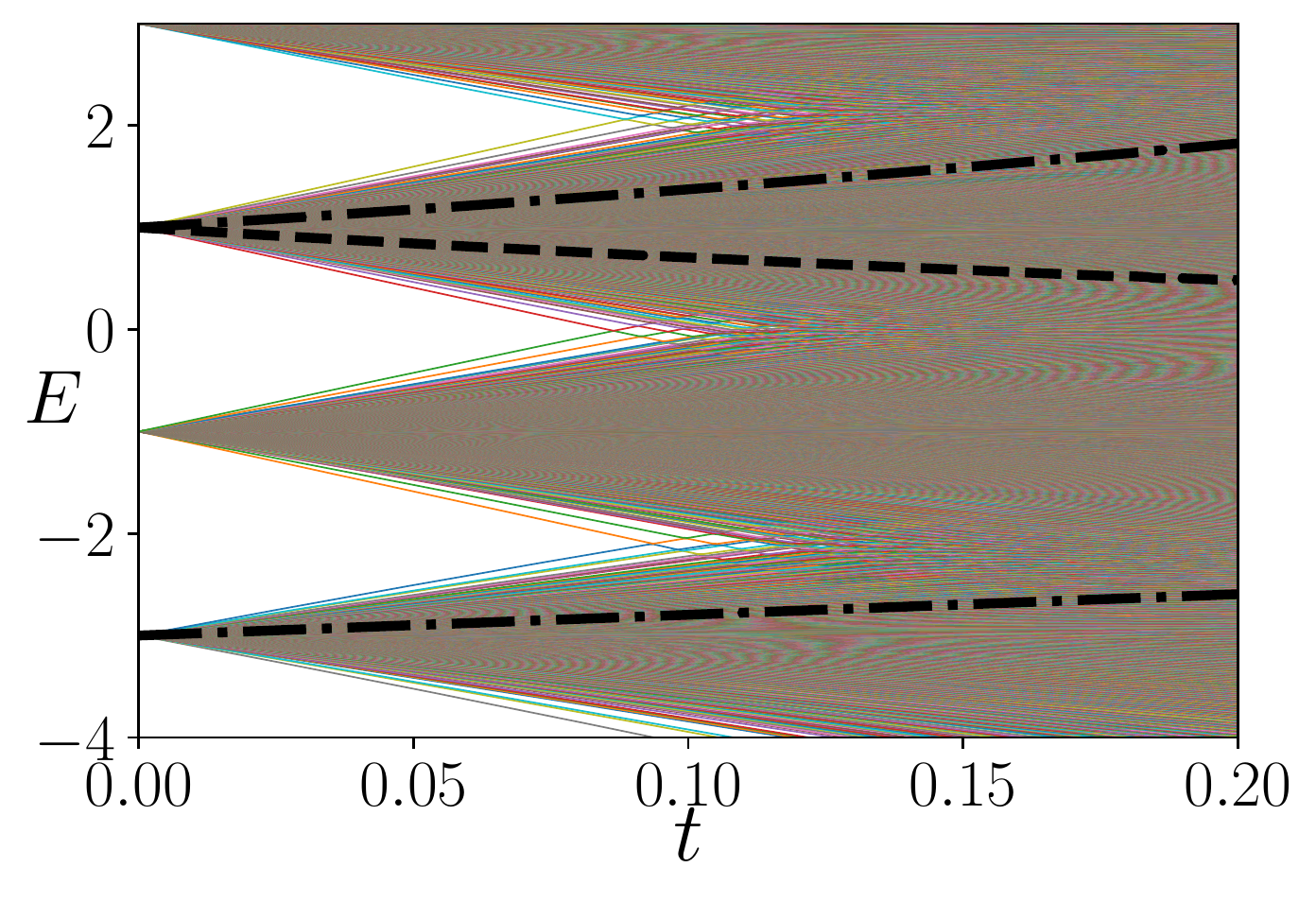}
	\caption{ The energy levels along the path ``0'''. The regularly spaced multiply degenerate levels at $h=t=0$ split when parameters increase along $t=3h$ line. The dashed line as well as two dash-dotted lines denote the position of antimagnon and two magnon excitations whose entanglement entropy is presented in Fig.~\ref{fig:new2}. As seen all three states lie in the middle, high density of states,  regime of the spectrum.}
	\label{fig:path0}
\end{figure}

\begin{figure}%
	\includegraphics[width=0.9\linewidth]{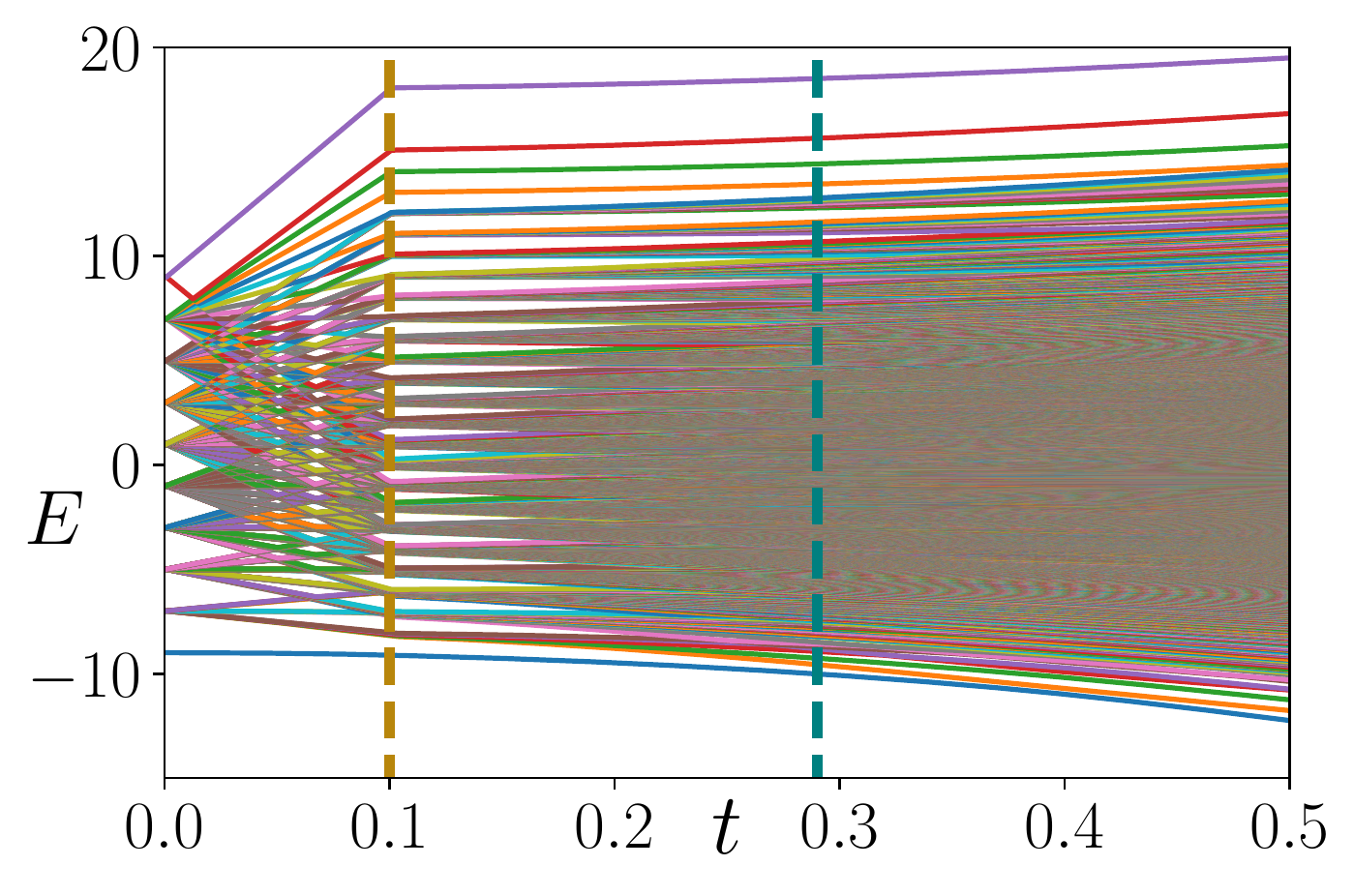}
	\caption{ The energy levels along the path-I. The regularly spaced levels split following the line $t=0.2h$ up to $t=0.1$. From there $h$ remains constant, $h=0.5$ while $t$ increases up to $t\approx 0.3$ which is the point where all the high density states have high entanglement entropy($\mathcal{S}$).}
	\label{fig:pathIa}
\end{figure} 
\subsection{Breaking of Adiabaticity on path `I' exanple}

The initial state $\vert S_n^k(t_0,h_0)\rangle$ is identified from the mapping between the Hamiltonian of  quantum Ising chain in both transverse and longitudinal fields (5) and the effective Hamiltonian (7) at $t_0,h_0=0.001$. The Hamiltonian, $H$, (5) is then diagonalized with the smaeq:kitll  parameter change of $\delta h=0.001$ and $\delta t=0.0002$ and a state $\vert E_n(t,h)\rangle$ is identified as the state which has a maximum overlap with $\vert S_n^k(t_0,h_0)\rangle$ among all the eigenstates of ${H}(t_0+\delta t,h_0+\delta t)$.

\begin{figure}%
	\includegraphics[width=\linewidth]{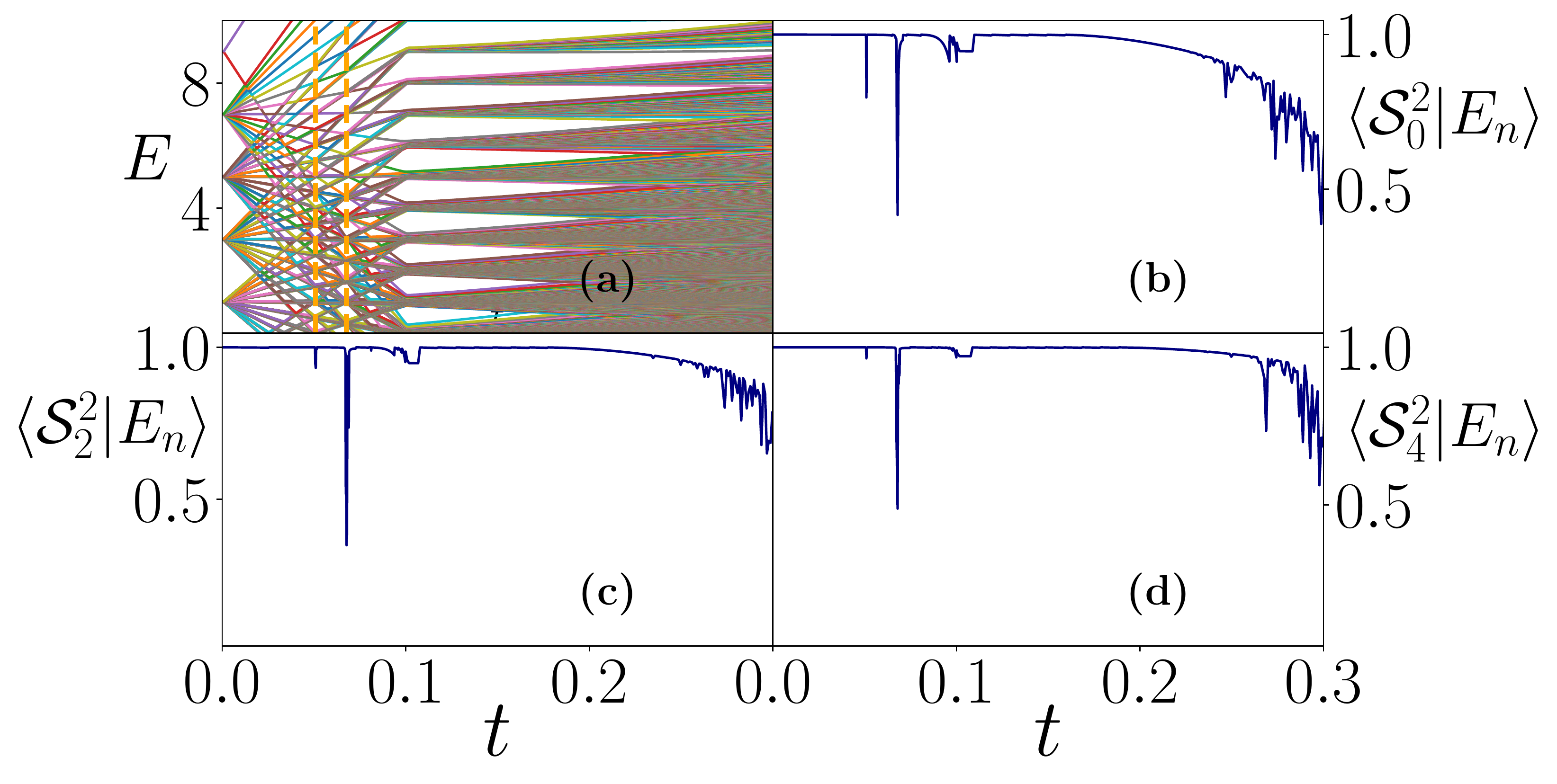}
	\caption{The overlaps along the path I of three antimagnon states. The sharp overlap drops at
	the beginning of the path, in almost integrable regions is due to multiple avoided crossings indicated by vertical dashed lines in the level diagram (top, left)}
	\label{figover}
\end{figure} 

Figure~\ref{figover} shows the exemplary overlaps along path I.

To adiabatically follow the suspected QMBS state $\vert S_n^k(t,h)\rangle$ is updated along the paths. But this is only done so, if for the next 10 diagonalizations (for a parameter change of $\Delta h=0.01$ and $\Delta t=0.002$) the eigenindex of the state $\vert E_n(t+\Delta t,h+\Delta h)\rangle$ does not change from its value at $\vert E_n(t,h)\rangle$. This procedure allows us to overcome problems encountered while passing
the avoided crossings (see also below).  The following protocol is continued until the final value of $h$ along a given path is reached. Later  we continue updating $t$ only with  $\delta t = 0.001$ while keeping $h$ constant.

\begin{figure}%
	\includegraphics[width=0.9\linewidth]{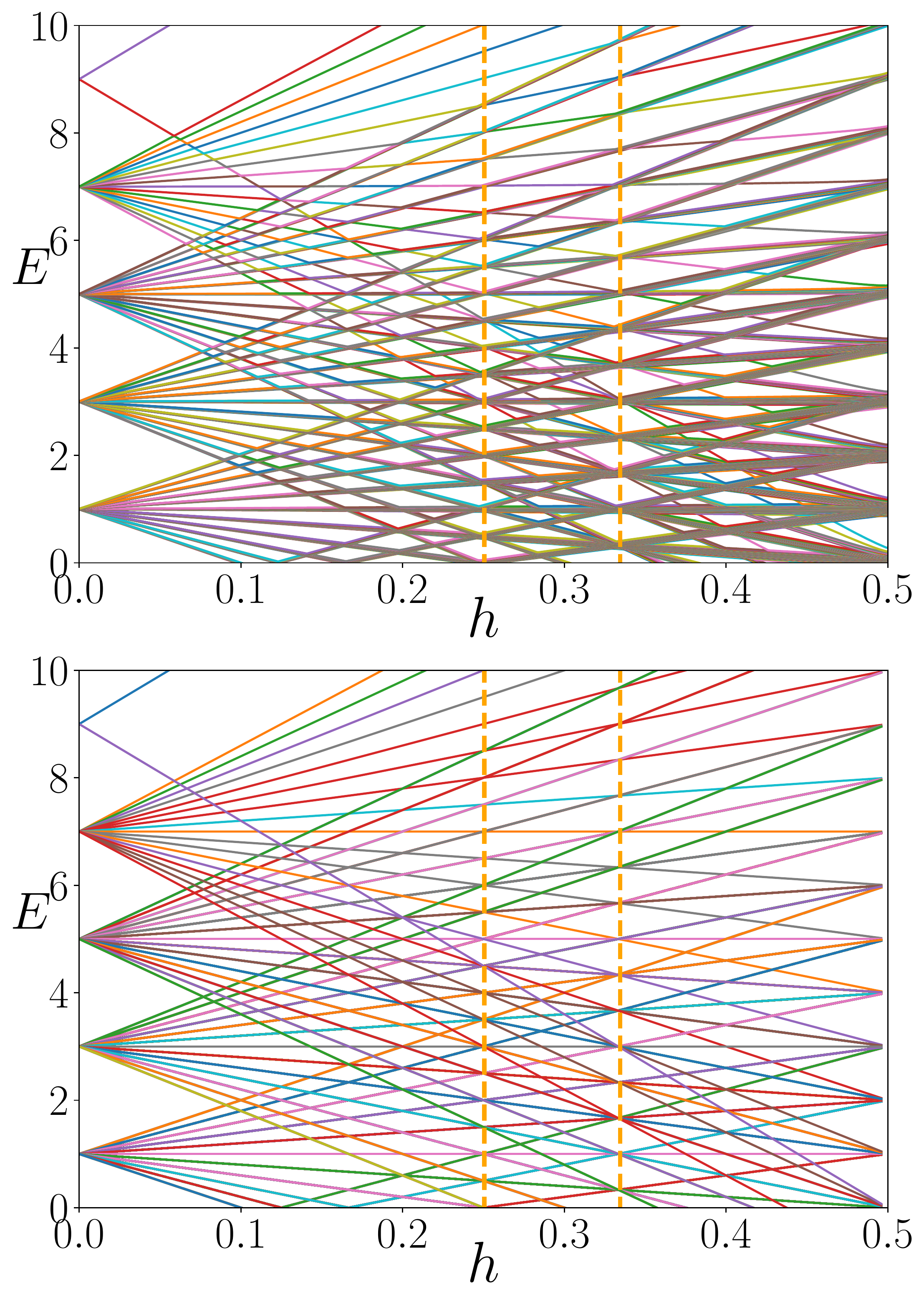}
	\caption{An enlarged image of Fig. \ref{fig:pathIa} -- top -- restricted to the parameters where the peaks  in entanglement entropy of the followed state and in $r$ value for the spectrum appear - compare Fig.~\ref{fig:contour}(c). Note that the range of $h$ corresponds via $t=0.2h$ relation defining path I to $t\in[0,0.1]$. The orange dashed lines represents the location of $t$ values for the peaks. The bottom plot shows the integrable system states energies obtained by putting $t=0$ in the Hamiltonian, \eqref{Isi}.
	\label{fig:en_lev}}
\end{figure} 

\begin{figure}%
	\includegraphics[width=0.9\linewidth]{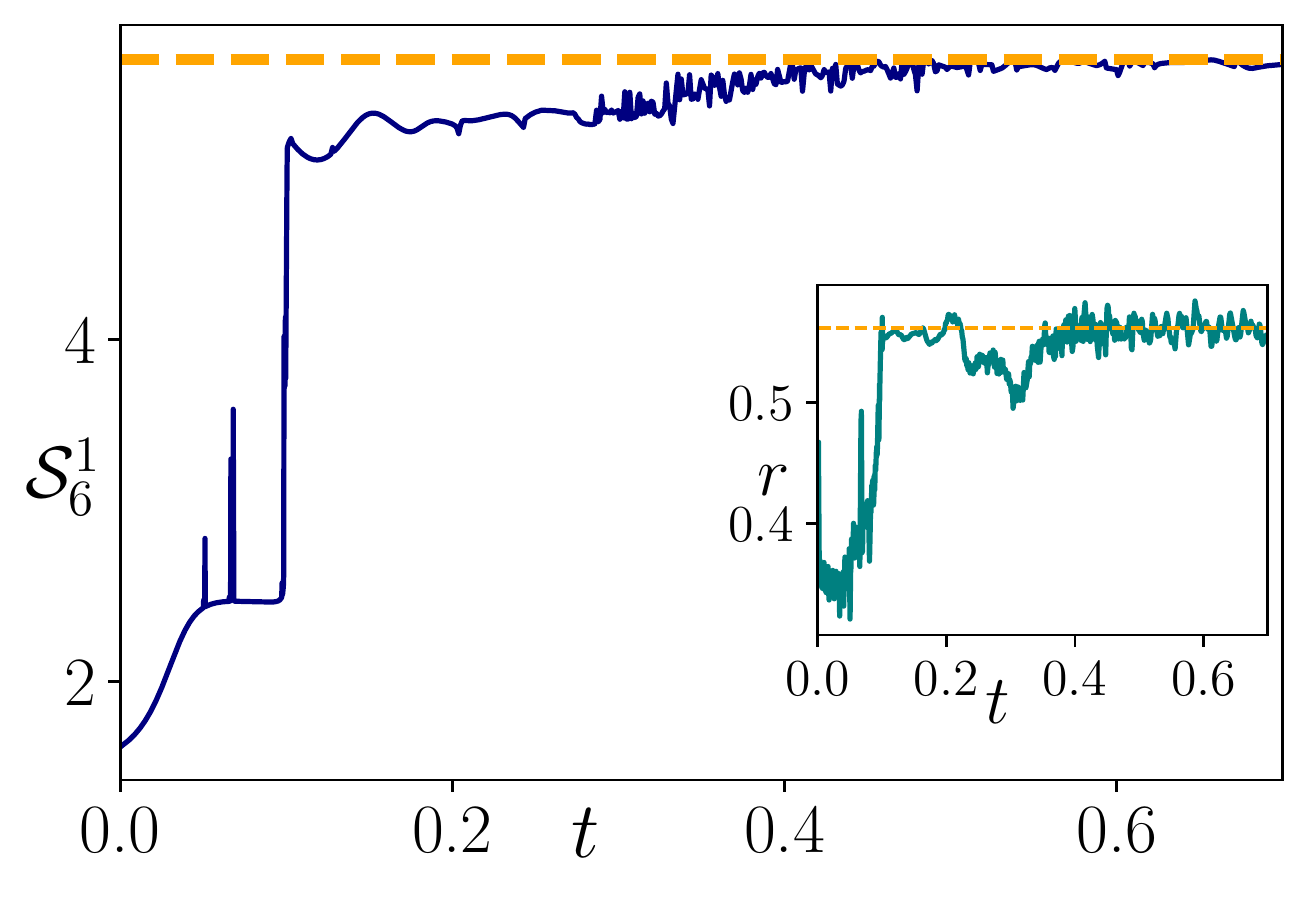}
	\caption{ Attempt to follow the magnon type state $S^1_6$ along path I. Once gap ratio reaches values close to GOE limit
	the entropy of the followed state rapidly and significantly increases indicating that the scarred character is lost.}
	\label{fig:magpa}
\end{figure} 

This procedure runs into difficulties when avoided crossings are met. In the regime corresponding to mixed regular and chaotic dynamics (as revealed statistically by the gap ratio, $r$, being between  Poissonian and GOE values) the avoided crossings are isolated. Such avoided crossings may be crossed diabatically  with care. Once we reach the irregular regime multiple avoided crossings typically arise and only those states can be traced that are almost decoupled in the process of changing the parameters. Such a situation is well known from chaotic, single particle, strongly scarred systems such as quantum billiards or hydrogen atom in a strong magnetic field (see e.g. \cite{Stockmann}). While in generalized time reversal invariant systems small avoided crossings are highly probable
\cite{Zakrzewski91,Zakrzewski93c} overlapping avoided crossings dissolve the scar character of the state unless a periodic orbit \cite{Heller84} or symmetry \cite{Delande87} reinforces this character after the crossing. While this intuition is based on a single particle systems, we expect the similar qualitative behavior for the possible preservation of QMBS.

Figure~\ref{fig:path0} presents the level dynamics along the path ``0'', i.e $t=3h$ line.
eq:kit
Fig.~\ref{fig:pathIa} shows the level dynamics for PATH I. At $t\in[0.0,0.1]$ the $r$ value is nearly Poissonian and all the identified QMBS states can be tracked. In fact, a simple structure of the tilted transverse Ising Hamiltonian (5) reproduced here
for clarity
\begin{equation}
	H = \sum_{i=1}^{L} \frac{\mu}{2} Z_i Z_{i+1} - t X_i   - h Z_i.
	\label{Isi}
\end{equation}
identifies that for $t,h\ll 1$ the states are ordered by number of dimers as given by  the operator $D=\sum_{i=1}^{L} Z_i Z_{i+1}$.
Once $h> 0$ those states form a manifold split by $\sum Z_i$ operator eigenvalues. Only at the crossings of the manifolds the term $ t \sum_i X_i $ mixes eigenstates of $Z_i$ leading to nontrivial avoided crossings, compare Fig.~\ref{fig:en_lev}. It is at these avoided crossings that sharp spikes visible in the entropy
$t$ scans along the path I appear. Here also the mean gap ratio shows the unusual peaks.

For larger  $t\in[0.1,0.3]$ the $r$ value is GOE-like indicating an ergodic character, in this region only the QMBS states built out of antimagnons are sustained indicating these states to be characteristically different from the other states. At $t > 0.3$ the levels become  strongly mixed and the Hilbert space splitting into characteristic bands disappears. Then all the tracked states loose their QMBS characteristics.

\subsection{Magnon States}

The scar states built out of magnon states which were located at the middle of the spectrum (in the region of a high density of states) tracked along path I were lost on entering the ergodic regime. This is in contrary to path ``0'' when both magnon and antimagnon excitations could be followed.  For the path I, only antimagnon states may be followed and linked to QMBS  in the ergodic regime. The example of our attempt to track a magnon state along path I is summarized in Fig.~\ref{fig:magpa}.

\end{document}